\documentclass[useAMS,usenatbib]{mn2e}
\RequirePackage{xspace}
\RequirePackage{amsmath}
\RequirePackage{amsfonts}
\RequirePackage{amssymb}

\def\ifundefined#1{\expandafter\ifx\csname#1\endcsname\relax}

\def\la{\mathrel{\hbox{\rlap{\hbox{\lower4pt\hbox{$\sim$}}}\hbox{$<$}}}}
\def\ga{\mathrel{\hbox{\rlap{\hbox{\lower4pt\hbox{$\sim$}}}\hbox{$>$}}}}

\newcommand{\be}{\begin{equation}}
\newcommand{\ee}{\end{equation}}

\newcommand{\bea}{\begin{eqnarray}}
\newcommand{\eea}{\end{eqnarray}}

\ifundefined{ensuremath}\def\ensuremath#1{\relax\ifmmode{#1}}
\else${#1}$\fi\else\relax\fi
\ifundefined{nuc}\def\nuc#1#2{\relax\ifmmode{}^{#1}{\protect\text{#2}}
\else${}^{#1}$#2\fi}\else\relax\fi
\newcommand{\gcm}{g~cm$^{-3}$\xspace}
\newcommand{\kmps}{\ensuremath{\text{km}~\text{s}^{-1}}\xspace}

\newcommand{\msol}{\ensuremath{{\text{M}_\odot}}\xspace}

\newcommand{\nni}{\ensuremath{\nuc{56}{Ni}}\xspace}

\def\ang{\ensuremath{\text{\AA}}\xspace}

\newcommand\phxO{\texttt{PHOENIX/1D}\xspace}

\newcommand{\gamray}{$\gamma$-ray\xspace}

\usepackage{graphicx}

\usepackage[T1]{fontenc} 
\usepackage{aecompl}
\usepackage{aas_macros}

\graphicspath{{./figs/}}
\def\altaffilmark#1{$^{#1}$}
\def\altaffiltext#1#2{$^{#1}$#2}
\newcounter{aaffilcoun}
\newcommand\theaaffil{\addtocounter{aaffilcoun}{1}\theaaffilcoun}
\newcounter{affilcoun}
\newcommand\theaffil{\addtocounter{affilcoun}{1}\theaffilcoun}
\title[Models for SN 2011fe]{Spectral Models for Early Time SN 2011fe Observations}

\author[Baron et al.]{E.~Baron\altaffilmark{\theaaffil,\theaaffil}, P.~Hoeflich\altaffilmark{\theaaffil},
  Brian Friesen\altaffilmark{1},
  M.~Sullivan\altaffilmark{\theaaffil},
  E.~Hsiao\altaffilmark{\theaaffil}, 
  R. S. Ellis\altaffilmark{\theaaffil},
\newauthor
  A. Gal-Yam\altaffilmark{\theaaffil},
  D. A. Howell\altaffilmark{\theaaffil,\theaaffil},
  P. E. Nugent\altaffilmark{\theaaffil},
  I.~Dominguez\altaffilmark{\theaaffil},
  K.~Krisciunas\altaffilmark{\theaaffil},
\newauthor
  M.~M.~Phillips\altaffilmark{5}, N.~Suntzeff\altaffilmark{13},
  L.~Wang\altaffilmark{13}, 
    R.~C.~Thomas\altaffilmark{11}\\
\altaffiltext{\theaffil}{Homer L.~Dodge Department of Physics and Astronomy,
  University of Oklahoma, 440 W.~Brooks, Rm 100, Norman, OK,
  73019-2061 USA}\\
\altaffiltext{\theaffil}{Hamburger Sternwarte, Gojenbergsweg 112, 21029
  Hamburg, Germany}\\
\altaffiltext{\theaffil}{Department of
  Physics, Florida State University, Tallahassee, FL 32306, USA}\\
\altaffiltext{\theaffil}{School of Physics and Astronomy, University of Southampton, Southampton SO17 1BJ}\\
\altaffiltext{\theaffil}{Las Campanas Observatory, Casilla 601, La Serena,
  Chile}\\
\altaffiltext{\theaffil}{Cahill Center for Astrophysics, California
  Institute of Technology, Pasadena, CA 91125, USA}\\
\altaffiltext{\theaffil}{Benoziyo Center for Astrophysics, Weizmann Institute
  of Science, 76100 Rehovot, Israel}\\ 
\altaffiltext{\theaffil}{Las Cumbres
  Observatory Global Telescope Network, Goleta, CA 93117, USA}\\
\altaffiltext{\theaffil}{Department of Physics, University of California,
  Santa Barbara, CA 93106-9530, USA} \\
\altaffiltext{\theaffil}{Computational
  Cosmology Center, Lawrence Berkeley National Laboratory, 1 Cyclotron
  Rd., Berkeley CA 94720, USA} \\
\altaffiltext{\theaffil}{Universidad de Granada
C/ Bajo de Huetor 24 Aptdo 3004, ES 18071, Granada, Spain}\\
\altaffiltext{\theaffil}{George P. and Cynthia
Woods Mitchell Institute for Fundamental Physics \& Astronomy,
Texas A \& M University, \\ Department of Physics \& Astronomy,
4242 TAMU, College Station, TX 77843, USA}
}

\begin{document}

\date{Accepted xxx Received xx; in original form xxx}

\pagerange{\pageref{firstpage}--\pageref{lastpage}} \pubyear{2015}

\maketitle

\label{firstpage}

\begin{abstract}
  We use observed UV through
  near IR 
  spectra 
  to examine whether SN 2011fe
  can be understood in the framework of Branch-normal SNe~Ia and to
  examine its individual
  peculiarities. As a  benchmark, we use a delayed-detonation model with
  a progenitor metallicity of $Z_\odot/20$. 
We study the sensitivity of
  features to variations in progenitor metallicity, the outer density
  profile, and the distribution of radioactive nickel. 
The effect of metallicity variations in the progenitor
have a relatively small effect on the synthetic spectra.
  We also find that the
  abundance stratification of SN~2011fe resembles closely that
    of a delayed detonation
  model with a transition density that has been fit to other 
  Branch-normal Type Ia supernovae. 
 At early times, the model photosphere is formed in
  material with velocities that are too high, indicating that the photosphere
recedes too slowly or that SN~2011fe has a 
lower specific energy in the outer $\approx 0.1 $ \msol than does the model.
 We  discuss several explanations for the discrepancies. Finally, we examine
 variations in both the spectral energy distribution and in the colors
 due to variations in the progenitor metallicity, which suggests that colors
 are  only weak
 indicators for the progenitor metallicity, in the particular explosion
 model that we have studied. We do find that the flux in the $U$ band
 is significantly higher at maximum light in the solar metallicity
 model than in the lower metallicity model and the lower metallicity
 model much better matches the observed spectrum.
\end{abstract}


\section{Introduction}\label{sec:Introduction}

Supernova PTF11kly/2011fe (henceforth SN~2011fe) was discovered by the
Palomar Transient Factory on Aug 23, 2011 in M101 only hours after
explosion \citep{nug_11fe_11}. This nearby, early discovered object
has been extremely well observed in all bands.  The Carnegie Supernova Project
obtained excellent spectroscopy in the infrared
\citep{hsiao11fe13}. Photometry has been obtained by \emph{SWIFT} in
the UV \citep{brown11fe12}, in the optical
\citep{RS11fe12,vinko11fe12,munari11fe13,factory11fe13}, and in the IR
\citep{matheson11fe12}. \citet{factory11fe13} presented a detailed
comparison of the photometric observations including a well calibrated
spectrophotometric time series. SN~2011fe has been
observed in the radio \citep{chomiuk11fe12}, in gamma-rays
\citep{isern11fe13a} and in the X-ray
\citep{liu11fe12}. From the spectra and photometry SN~2011fe is about
as ordinary a SN~Ia as there could be. Due to the early
discovery and close proximity, several groups have drawn conclusions
about the environment of SN~2011fe. Using non-detections and the very
early observed points \citet{bloom11fe12} were able to constrain the
radius of the primary star, concluding that it must be a compact
object (white dwarf or neutron star).  Using archival \emph{HST}
images \citet{weidong_last} were able to rule out  luminous red giant
and helium star companions.  Using radio data, constraints have been
placed on the progenitor environment which have been interpreted as
ruling out the single degenerate scenario \citep{chomiuk11fe12};
however, the winds blown during the progenitor formation could
naturally produce a low density environment that do not necessarily
require a degenerate companion \citep[see also][]{horesh11fe12}.

Some spectral  modeling of the optical spectra of SN~2011fe has 
been presented \citep{roepke11fe12,dessart11fe14}. Here, we want to compare and
contrast SN~2011fe 
to a model which reproduces the spectra of a Branch-normal supernova \citep{sandage90N96}. We use a
model because this allows us to translate spectral similarities and
differences into
physical space, for example, to determine the relevant mass layer
involved in the differences. We have specifically chosen a generic
DD model for Branch-normal supernova and not tuned the model for this
specific supernova.
The goal of this study is to show that SN~2011fe is close to a
Branch-normal supernova and to 
evaluate differences and discuss possible physical causes.

 This study is based on the combined UV and optical \textit{HST}  and IR Gemini spectra.
The infrared spectra that we study have been already were presented in 
\citet{hsiao11fe13}.  The UV+optical observations were also presented
and studied in previous work
\citep{foley12a,foley12b,FK13,mazz11fe14}. Here, we present a
  detailed comparison quantitative synthetic spectroscopy with a
  number of epochs of SN~2011fe, all of which include coverage from
  the UV through the IR.

\section{Spectral calculations and explosion model}\label{sec:Model}

We performed spectral calculations using the multi-purpose stellar
atmospheres program \phxO\ {version \tt 16}
\citep{hbjcam99,bhpar298,hbapara97,phhnovetal97,phhnovfe296}. Version
16 incorporates many changes over previous versions used for supernova
modeling \citep{bbh07,bbbh06} including many more species in the
equation of state (83 versus 40), twice as many atomic lines, many
more species treated in full NLTE, and an improved equation of state.
\phxO solves the radiative transfer equation along characteristic rays
in spherical symmetry including all special relativistic effects.  The
non-LTE (NLTE) rate equations for many ionization states are solved,
including the effects of ionization due to non-thermal electrons from
the \gamray{s} produced by the  radiative decay of $^{56}$Ni, which
is synthesized in the
supernova explosion.  The atoms and ions calculated in NLTE are
  He~I--II, C~I--III, O~I--III, Ne~I, Na~I--II, Mg~I--III,
  Si~I--III, S~I--III, Ca~II, Ti~II, Cr~I--III, Mn~I--III, Fe~I--III,
  Co~I--III, and Ni~I--III. 
 These are all the
elements whose features make important contributions to the observed
spectral features in SNe~Ia. 

Each model atom includes primary NLTE transitions, which are used to
calculate the level populations and opacity, and weaker secondary LTE
transitions which are included in the opacity and implicitly
affect the rate equations via their effect on the solution to the
transport equation \citep{hbjcam99}.  In addition to the NLTE
transitions, all other LTE line opacities for atomic species not
treated in NLTE are treated with the equivalent two-level atom source
function, using a thermalization parameter, $\alpha =0.10$
\citep{snefe296}.  The 
atmospheres are iterated to energy balance in the co-moving frame;
while we neglect the explicit effects of time dependence in the
radiation transport equation, we do implicitly include these effects,
via explicitly including $p\,dV$ work and the rate of gamma-ray
deposition in the generalized 
equation of radiative equilibrium and in the rate equations for the
NLTE populations.

The outer boundary condition is the total bolometric luminosity in the
observer's frame. The inner boundary condition is that the flux at the
innermost zone ($v=700$~\kmps) is given by the diffusion
equation. Converged models required 256 optical depth points to
correctly obtain the Si~II $\lambda 6355$ profile.

\begin{figure}
\includegraphics[scale=0.4]{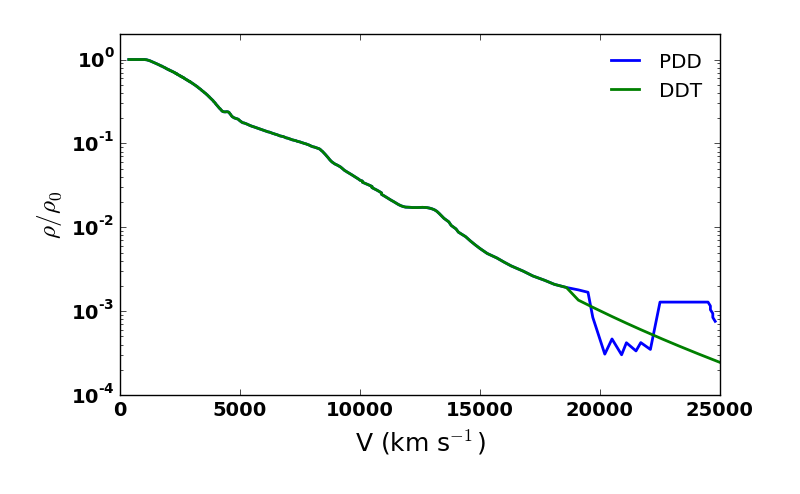}
\caption{The density profile of the prompt DDT 
  and the PDD that were compared. The value of $\rho_0$ at maximum
  light is about $2\times 10^{-12}$~\gcm.}
\label{fig:rho_orig_rho_new}
\end{figure}

For our analysis, we use a delayed-detonation model which 
reproduces the light curves and spectra for Branch-normal supernovae
\citep{hofdd+mol95,HGFS99by02,hoef_review06,gerardy03du04}. \citet{gerardy03du04}
compared a very similar model to SN~2003du for the entire range of optical to 3800 \AA\ to 2$ \mu$m.
  The models start from a
C/O white dwarf taken 
from the core of an evolved 5\msol main sequence star.  Through
accretion, this core approaches the Chandrasekhar limit.  An explosion
begins spontaneously when the core has a central density of $2.0
\times 10^9$~\gcm and a mass close to $1.37$~\msol
\citep{h02}. The transition from deflagration  to detonation is
triggered at a density of $2.3 \times 10^{7}$~\gcm . We considered two
modes for the delayed-detonation transition: one with a direct transition
during the deflagration phase, the other after
a mild pulsation  which formed an envelope of $10^{-2}$~\msol. 
The resulting density structures are shown in  Fig.~\ref{fig:rho_orig_rho_new}.
We considered initial metallicities $Z$ of $Z_\odot$ and $Z_\odot/20$.
Here, $Z$ is defined as the iron abundance relative to solar. We take into
account the smaller variation of the elements such as Ne and O
compared to the Fe-group elements 
 \citep{argast_rev01}. 
The former dominates the metallicity effect on nuclear burning, whereas
the latter sets the floor for Fe-group elements \citep{hwt98,hnuw00}. 
 
\section{Results}\label{sec:results}

\subsection{General Properties}

\begin{figure}
\includegraphics[scale=0.4]{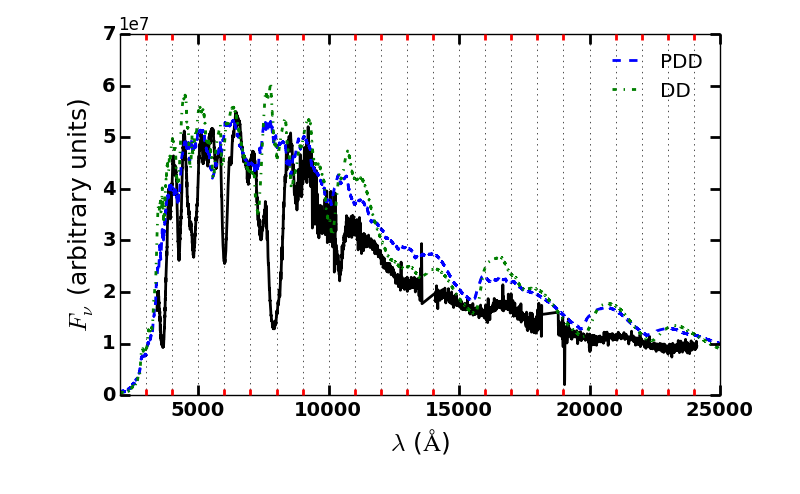}
\caption{PDD vs DD at Day 02. The spectra of the PDD model (dashed line) and the DD
  model (dot-dashed line) are compared to the observations (solid line).}
\label{fig:pdd_vs_dd_day02}
\end{figure}

 We calculated both a classical delayed detonation (DD) model and a
 pulsating delayed detonation (PDD) model motivated by
 the report of  \citet{jerod11fe12} 
of two distinct high and low velocity components in the Si, O, and C lines.
 In Figure~\ref{fig:pdd_vs_dd_day02}, we show the comparison between
 the DD model, the  PDD model, and the observations at day 2.
While neither model is strongly preferred over the other, the features
in the 
PDD model are more washed out, a trend which continues into  later epochs
and it  indicates that the dense shell is
certainly not desirable, thus we choose to use the
standard DD model for this study. Our PDD model and DDT model differ
somewhat from those of \citet{dessart11fe14}, but we find that the
variation in the colors, specifically the flux in the $U-$band is due
to metallicity of the progenitor.

We find a preference between
models with solar metallicity and  $Z_\odot/20$. There are not strong
differences at the earliest times, but rather at maximum light. The solar metallicity
progenitor  produces  somewhat redder 
 spectra in the bluest bands and we adopted a
 progenitor 
 metallicity with $Z_\odot/20$ as the fiducial model. We discuss the
 differences in \S~\ref{subsec:progmet}.  
  
\subsection{Benchmarking SN~2011fe}
\label{sec:res:bench}

Having probed the primordial metallicity and explosion class, we use a
DD model whose calculated light curve gives results in line
  with the observed class of Branch-normal supernovae as a benchmark for SN~2011fe.  With
time, spectra reveal increasingly deeper layers. The spectral sequence provides
a key probe of the layers for similarities and differences between the
model and the observations. Therefore,
throughout our discussion we will identify the layers by their mass
coordinate measured from the outside --- more precisely, we report
the location of the line forming region of the Si~II $\lambda6355$
line. The velocity of both the observed and synthetic line is measured
the same way, by determining the wavelength of the blueshifted minimum.
Note that the UV spectra always probe the very outer layers
\citep[see Fig.~11 in][]{hofdd+mol95}.

We consider epochs up to maximum light because the corresponding
``photospheric regions'' undergo partial burning.  
  Details of the spectral features are rather
sensitive to small variations in brightness and non-thermal excitation
\citep{b01ay}.

\begin{figure}
\includegraphics[scale=0.4]{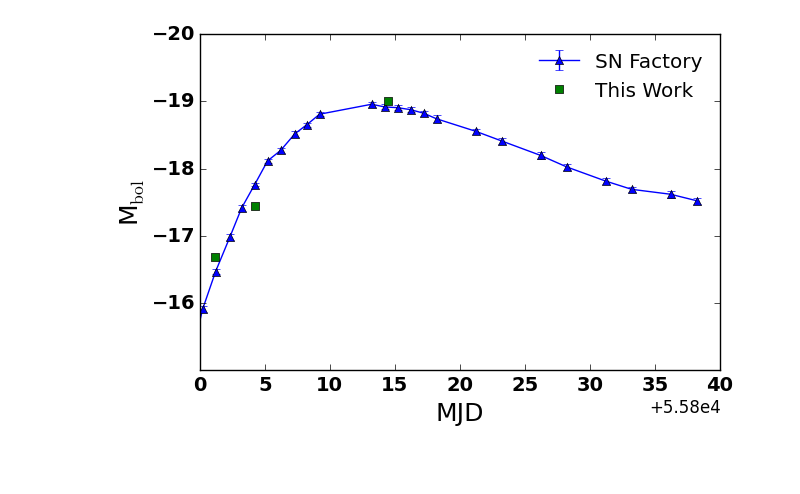}
\caption{The bolometric light curve from these calculations compared
  to the bolometric light curve inferred by \citet{factory11fe13}.}
\label{fig:bol_lc}
\end{figure}

\paragraph{Aug 25/Day 2}

\begin{figure}
\includegraphics[scale=0.4]{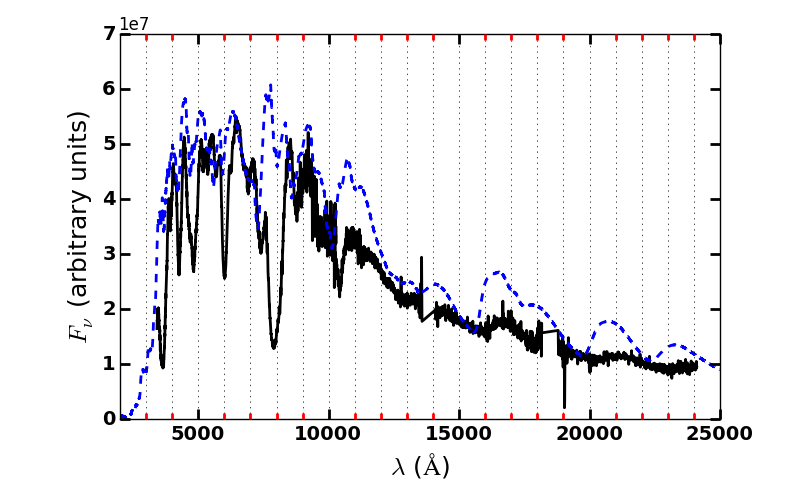}
\caption{NLTE synthetic spectrum on day 02 (dashed line) is compared to the optical
  spectrum obtained at Lick on Aug 25, 2011 and the IR spectrum
  obtained at Gemini North on Aug 25, 2011 (solid line).}
\label{fig:day02l4_nlte}
\end{figure}

The earliest spectrum of SN~2011fe was obtained on Aug. 25th, about 2
days after the explosion, which probes the outer $5 \times 10^{-3}
M_\odot$.  The spectral features are quite sensitive to the
temperature and
excitation, we show the NLTE spectra in
Figure~\ref{fig:day02l4_nlte}. We calculated an LTE model with the
same parameters and  the LTE model is much too bright in
the IR and the features are weak, compared to the NLTE model.
 The 
figures show that almost all the lines present in the observed
spectrum are also in the synthetic spectrum. While many of the line
strengths are well reproduced by the model, the Si~II $\lambda 6355$
line is much too weak and the Si~II $\lambda 5970$ line is about the
right strength, but is too fast. The LTE model over-predicts the IR
flux and the IR features are weak, whereas the NLTE model roughly gets
the IR continuum correct, but the IR features are too strong and
broad.
A common
feature of both the LTE and NLTE spectra is a too large
Doppler shift of all the absorption features of intermediate mass
elements both in the optical and IR, including the Ca II IR-triplet
and the O~I $\lambda 7773.4$ line.  As
mentioned above, details of the line-features are very sensitive to
excitation and temperature effects.  In particular, a higher
Doppler-shift may be produced by an excessive emission component seen
for example in Mg~II $\lambda 10926$ or O~I $\lambda 7773.4$.  However,
even in case of LTE, the trend is confirmed.  The smaller Doppler shifts in
SN~2011fe compared to the delayed detonation model for a Branch-normal supernova may be attributed to a
lower specific energy or photosphere that recedes more quickly.

\begin{figure}
\includegraphics[scale=0.4]{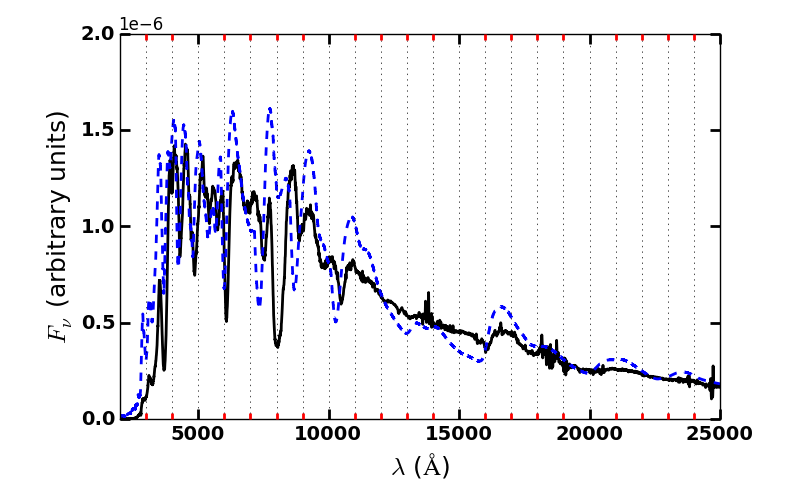}
\caption{NLTE synthetic spectrum on day 05 (dashed line) is compared to the UV/optical
  spectrum obtained by \emph{HST} on  on Aug 28, 2011 (MJD~55801.12)
  and the IR spectrum obtained at Gemini North on Aug 28, 2011 (solid line).}
\label{fig:day5l7}
\end{figure}

\paragraph{Aug 28/Day 5}
Figure~\ref{fig:day5l7} shows the comparison  to the combined
\emph{HST}+optical+IR spectrum obtained on MJD~55801.12 
 which probes the outer $3 \times 10^{-2} M_\odot$.
 The
overall spectral shape from UV through optical is quite well
reproduced. The far UV is a bit high, but the quality of the fit from
$0.2-2.5\ \mu$m is quite good.  the Si~II $\lambda
6355$ line is about the right strength, the Si~II $\lambda
5970$ line is too strong and fast by about 4,000~\kmps, as is the Ca
IR triplet. Actually, the the Si~II $\lambda
5970$ line is low by about 20 percent, but it does not recover to the
blue as the observed spectrum does, due to blending with other lines.
In
general the spectrum is somewhat too fast. 

\paragraph{Aug 31/Day 8}

\begin{figure}
\includegraphics[scale=0.4]{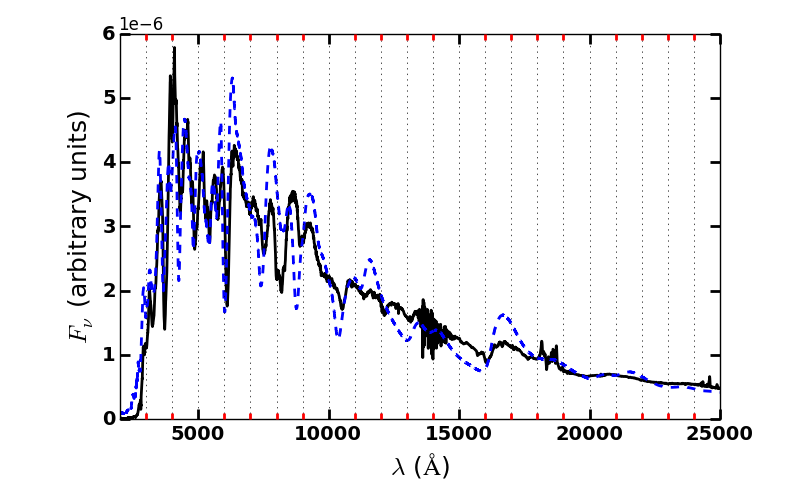}
\caption{NLTE synthetic spectrum on day 08  (dashed line) is compared to the UV/optical
  spectrum obtained by \emph{HST} on  on Aug 31, 2011 (MJD~55804.25)
  and the IR spectrum obtained at Gemini North on Aug 31, 2011 (solid line).}
\label{fig:day8l4}
\end{figure}

Figure~\ref{fig:day8l4} shows the comparison to the combined
\emph{HST}+optical+IR spectrum obtained on MJD~55804.25
 which probes the outer $5 \times 10^{-2} M_\odot$.
 Not only is
the overall spectral shape from UV through optical quite well
reproduced, but also the relative strength of the far UV closely
matches the observations.
However, the iron and silicon 
feature around 4000~\ang is too weak. The strength of
Ca II H+K is well reproduced. The calcium IR
triplet is now too narrow, and not strong enough in absorption. The
Si~II $\lambda 
6355$ line is well fit in absorption strength, but it is too fast by
about 4,000 \kmps
and the emission peak is a about 20\% too high, indicating that the model at
this phase is too extended. The Si~II $\lambda
5970$ line is now also well fit in absorption. The Mg~II $\lambda 9226$
line is much too strong. The Mg~II $\lambda 10926$
feature is too strong as are most of the rest of the features further
to the red in the IR. The O~I $\lambda 7773.4$ line is prominent, but
too strong and too broad. The Si~II $\lambda 11714.87$ feature is
prominent in the synthetic spectrum, but much weaker in the
observed spectrum.

\paragraph{Sep 10/Day 18}

\begin{figure}
\includegraphics[scale=0.4]{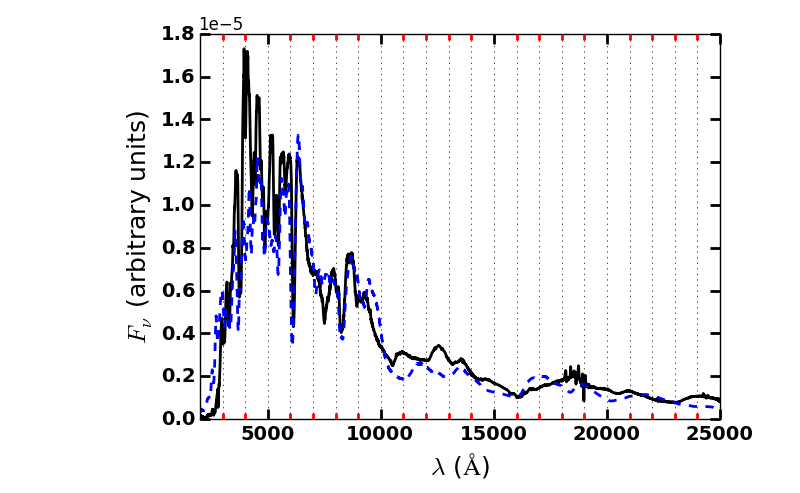}
\caption{NLTE synthetic spectrum on day 18 (dashed line) is compared to the UV/optical
  spectrum obtained by \emph{HST} on  on Sep 10, 2011 (MJD~55814.43)
  and the IR spectrum obtained at Gemini North on Sep 10, 2011 (solid line).}
\label{fig:day18l3}
\end{figure}

In Fig. ~\ref{fig:day18l3}, we show the spectrum obtained on
MJD~55814.43 which which is close to maximum light
\citep{nug_11fe_11,factory11fe13} which probes the outer $0.5
M_\odot$.  For reference, \citet{factory11fe13} find
$t_{B_\mathrm{max}}$ on MJD~55814.51.  The
observed spectrum of SN~2011fe closely resembles the synthetic
spectrum for a Branch-normal supernova both with respect to the line
strength and the Doppler shifts of individual lines.  Overall the
Doppler shifts are consistent, but the model is slightly too cool,
resulting in a reduced flux in the blue. This could be due either to
strong line blanketing in the blue, which pushes flux to the red or
because the opacities due to the iron group elements decrease in the
blue as the temperature decreases. The complex formation of the
observed spectrum was studied in \citet{bongard08}. The Si~II $\lambda 6355$
line is now in good agreement in both absorption and emission
strength, indicating that the model at this phase has about the
correct velocity extension.  The Si~II$\lambda 5970$ line now is also
well fit in both absorption and emission (modulo the too low pseudo
continuum).    The Ca
II IR triplet is too weak. The Mg~II $\lambda 9226$ and 
Mg~II $\lambda 10926$ features are too strong, as
are most of the rest of the features further to the red in the IR.

\section{Discussions and Conclusion}
\label{sec:discussion}

\begin{figure}
\includegraphics[scale=0.4]{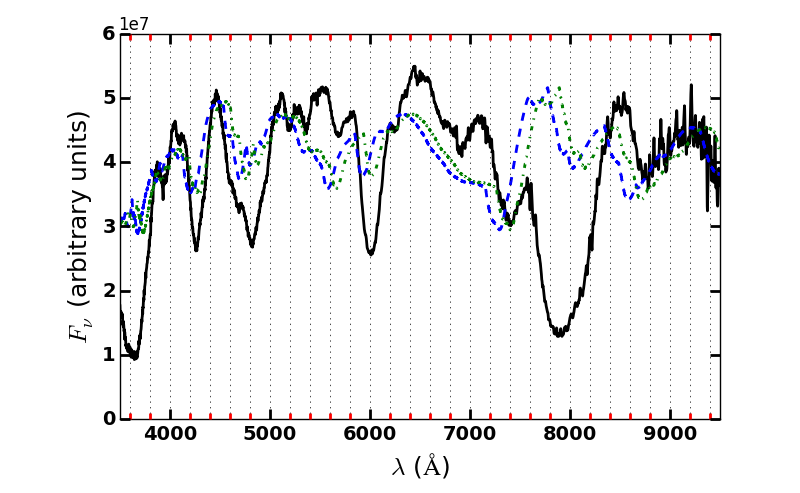}
\caption{Optical blow up of Aug 25 spectrum (solid line) shown in
  Fig~\ref{fig:day02l4_nlte} where the synthetic spectra are shown with
  (dot-dashed line)
  and without (dashed line) a ``redshift'' of $z=0.015$.}
\label{fig:day02l4_nlte_opt_wred}
\end{figure}

Figure~\ref{fig:day02l4_nlte_opt_wred}
shows the optical spectra of the Aug 25 spectrum we have modeled, along with the
synthetic spectra shifted by a ``redshift'' $z =
0.015$. Similar experiments give ``redshift values of
$0.015,0.015,0.010$, for Aug 28, Aug 31, and Sep 10,
respectively. This shows that the recession 
of the pseudo photosphere in the  hydro model with time is slower than in
the observations. 
We can estimate the amount of the shift by
comparing the velocity of the absorption minimum of Si~II $\lambda
6355$ line at each epoch. Table~\ref{tab:vphot} shows the measured
minimum velocity for each epoch. Since the velocity is close at
maximum light we will consider the ``outer part'' of the model to be
at 
velocities $v \ge 11,300$~\kmps.

\begin{table}
\begin{tabular}{rrr}
\multicolumn{3}{c}{Si~II $\lambda 6355$ Velocities}\\
\hline\hline
Epoch (Day)   & $V_\mathrm{Si II}$  & $V_\mathrm{Si II}$  \\
(since explosion)&(observed)&(synthetic)\\\hline
2&16,300&20,700\\
5&13,000&17,300\\
8&11,300&16,000\\
18&9,700&10,100\\\hline
\end{tabular}
\caption{Velocities of the absorption minimum of Si~II $\lambda
  6355$ feature at each epoch in \kmps. }
\label{tab:vphot}
\end{table}  

\begin{figure}
\includegraphics[scale=0.4]{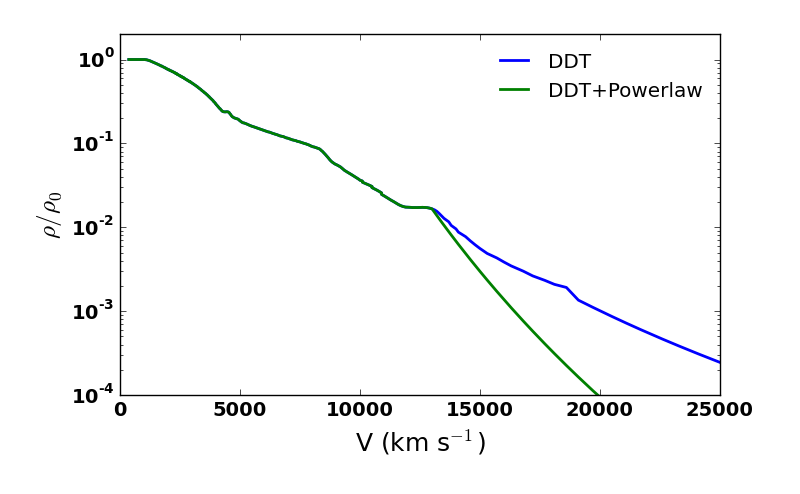}
\caption{The original density profile of the hydro model is compared
  to a modified profile where the density is forced to follow a
  powerlaw $\rho \propto (v/v_\mathrm{cut})^{-n}$, for velocities $v >
v_\mathrm{cut}$,  where $v_\mathrm{cut} = 12,000$~\kmps, and
$n = 12$.}
\label{fig:rho_orig_rho_cut_n12}
\end{figure}

\begin{figure}
\includegraphics[scale=0.4]{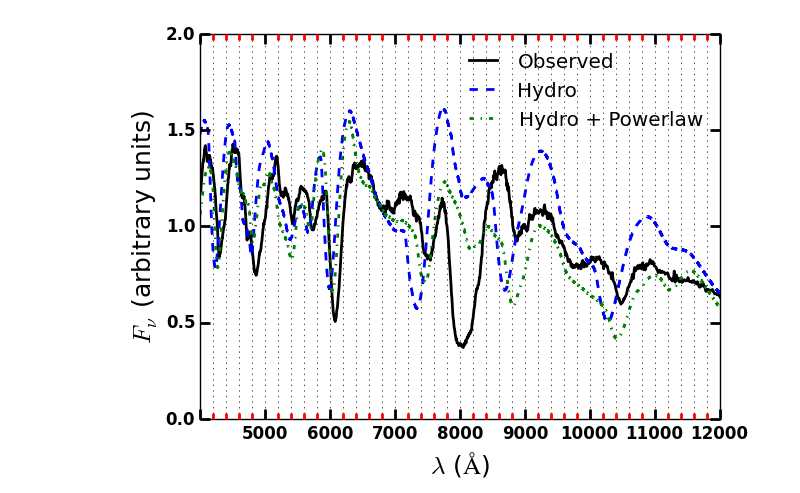}
\caption{NLTE spectra day 05 original hydro model (dashed line) compared to
  model hydro model with  density profile $\rho \propto v^{-12}$ for $v >
  12,000$~\kmps (dot-dashed line). The observed spectrum is shown with
a solid line.}
\label{fig:day05_n12_vcut12_fnlte}
\end{figure}

We can reduce the effective velocity extension by steepening the
density profile beyond the
photosphere. This can be produced by an outwardly increasing specific
energy with mass element. While steeper density profiles violate
energy conservation from nuclear burning and make the models
hydrodynamically inconsistent, we perform an empirical
exercise to examine its effect on the synthetic spectra, being careful
not to generalize the results too much.
 Figure~\ref{fig:rho_orig_rho_cut_n12} shows the modified
density profile, obtained by forcing the density to follow a power-law
$\rho \propto (v/v_\mathrm{cut})^{-n}$, for velocities $v >
v_\mathrm{cut}$,  where we took $v_\mathrm{cut} = 12,000$~\kmps, and
$n = 12$ \citep{branchcomp105,branch_pre07,branchcomp206,branch_post08,branchcomp509,doull11}. 
Figure~\ref{fig:day05_n12_vcut12_fnlte} shows a comparison of the Aug
28/Day 5 spectrum with and without the density profile
modification. The effects of the density modification are largest in
the red and smaller in the blue. The Si~II$\lambda
6355$ is somewhat slower and the line profile is clearly narrower, but
the agreement is not significantly better. However, in the red, the shape of
the O~I $\lambda 7773.4$ line is much better and that continues on to
the Ca~II IR triplet, the  Mg~II $\lambda 9226$ feature, and the Mg~II
$\lambda 10926$ feature. In the modified density structure the Si~II
$\lambda 11714.87$ becomes more pronounced. Thus, while steepening the
density profile 
does not significantly improve the agreement, it does have some benefit in
moving the absorptions of the redder lines to the correct velocity.

A full study of the possible reasons for lower observed photospheric
velocities is beyond the scope of the current study. Some possible
reasons for lower velocities can be understood within the framework of
spherical DD models. For a wide range of model parameters the overall
density structures are very similar \citep{hmk93b}. A possible reason
for the lower observed photospheric velocity may be a faster receding
photosphere in mass. To first order for one to two weeks past maximum
light in Branch-normal SNe~Ia, the opacities remain high and the
photosphere recedes in mass due to geometrical dilution of the
expanding envelope \citep{hmk93b}. In order to increase the recession
rate of the pseudo-photosphere we need a lower kinetic energy. In
classical DD models, most of the matter undergoes burning. Possible
ways to reduce the kinetic energy include: higher binding energy of
the white dwarf and thus, higher central densities; less nuclear
energy production due to a smaller C/O ratio, by, for example, a
larger main sequence mass; less heating in the outer layers by \nni
by shifting its distribution to the central layers. We are not
signaling out one particular cause and to do so is beyond the scope of
this work. We note that \citet{b01ay} suggested that a higher value
of C/O could explain the high velocities seen in SN~2001ay, whereas
\citet{ohlmann14} found that varying the C/O ratio alone in a fixed
white dwarf did not have dramatic effects.

PDD models with a large amount of unburned material will
reduce the explosion energy proportional to the amount of unburned material,
and may  lower the velocities of elements of incomplete burning.  However,
strong pulsations produce an outer, high velocity layer as observed
in, for example,  SN~1990N and similar SNe~Ia \citep{quimby05cg06}.
An alternative explanation is significantly lower mass progenitor than the
Chandrasekhar mass which,
would be less luminous and inconsistent with the light curve.

\subsection{Progenitor Metallicity}
\label{subsec:progmet}

\begin{figure}
\includegraphics[scale=0.45]{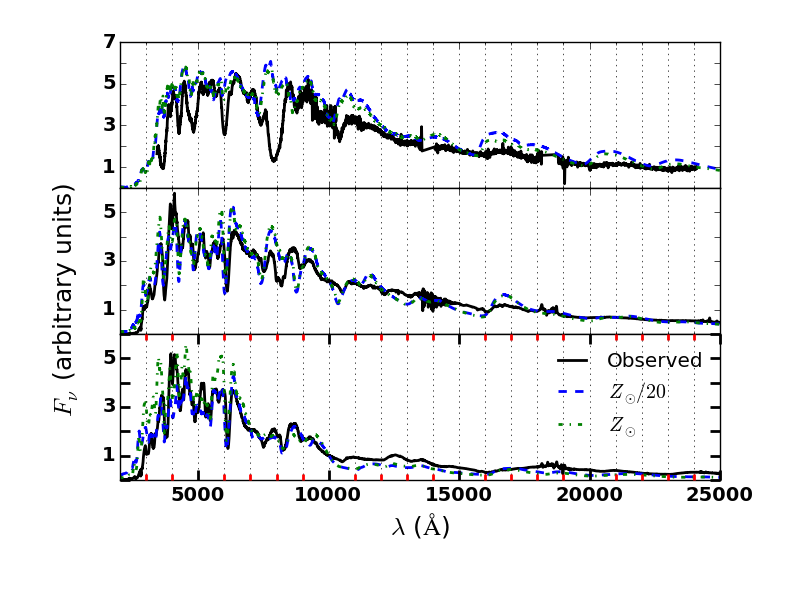}
\caption{The synthetic spectra of the $Z = Z_\odot/20$ (dashed line) and
  the $Z = Z_\odot/20$ (dot-dashed line) progenitor model compared to
  observations (solid line) for days Aug 25 (top panel), Aug
  31 (middle panel), and Sep 10 (bottom panel).}
\label{fig:metal_comp}
\end{figure}

\begin{figure}
\includegraphics[scale=0.4]{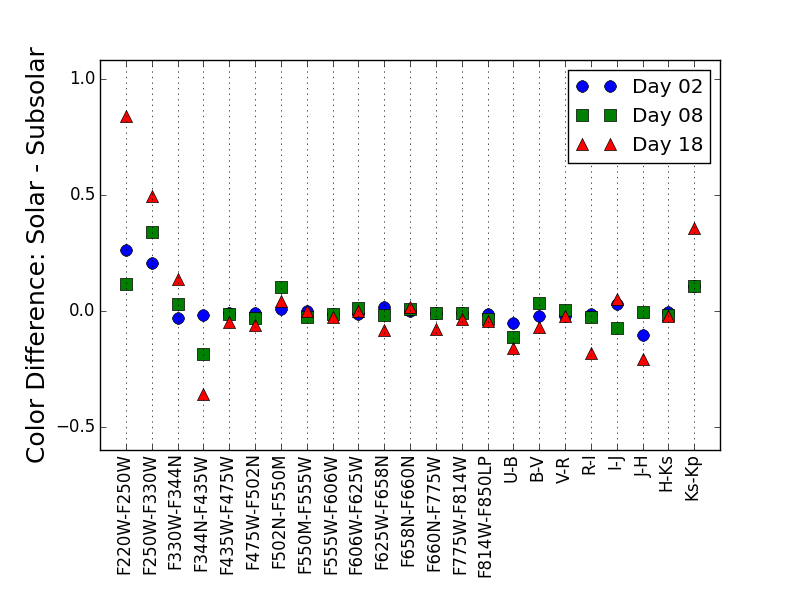}
\caption{The difference in 23 synthetic colors for $Z = Z_\odot$ the
  $Z = Z_\odot/20$ progenitor model compared for days Aug 25, Aug 31,
  and Sep 10. The colors are plotted from bluest to reddest filters
  and the difference plotted is, for example,
  $(B-V)_{Z_\odot} -(B-V)_{Z_\odot/20}$. The filters used are in order
  from blue to red, HST/ACS-HRC: F220W, F250W, F330W, F344N, F435W,
  F475W, F502N, F550M, F555W, F606W, F625W, F658N, F660N, F775W,
  F814W, F850LP, F892N; UBVRI+2MASS: U,B,V,R,I,J,H,K. We did not calculate
  colors across the filter sets, so the F892N-U color is not plotted.}
\label{fig:color_scatter}
\end{figure}

Figure~\ref{fig:metal_comp} compares the synthetic spectra of the
fiducial progenitor model with $Z = Z_\odot/20$ to that of progenitor
with solar metallicity. The other explosion parameters were the same
for both models. By eye, prior to maximum light, there is not a clear
preference for 
either model, which is somewhat surprising since na\"ively one expects
progenitor metallicity to play a role at particularly early
times. At maximum light a strong discrepancy appears,
where the solar metallicity progenitor appears too blue or more
accurately, the flux is much higher in the $U-$band.
 To attempt to quantify this we
calculated synthetic photometry on both the observations and the
models and compared the colors. This procedure is subject to
systematic errors since it assumes that the relative flux of the
observations is very accurate, so that the synthetic colors calculated
from the observed spectra are meaningful. Given this caveat, we did not
find a clear preference for either progenitor metallicity, based
\textit{solely}  on
colors. We examined 
the 25 filters from \textit{HST} and the Johnson set $UBVIRJHK_s$
and looked at all combinations of neighboring blue filter - red
filter. The lack of clear trends can be seen in the 
bluest filters F220W - 
 F250W, the models are too red at day 02, and too blue at days 08 and
 18, in the   F250W - F330W filters, the models are significantly too
 blue at day 02, but then only slightly too blue at days 08 and 18.
 The discrepancy is far smaller in the
F330W- F344N  and F344N - F435W colors,  with the same trend and then
changes in the F435W - F475 color where the models are too blue at day
02, too red at day 08 and then the solar model is slightly too blue
and the sub-solar model slightly too red at day 18. 
Figure~\ref{fig:color_scatter} shows the differences in the
twenty-three different colors at the three different epochs
studied. The trend for the solar metallicity model to become much
redder in the bluest bands 
at maximum light is evident, but it is difficult to discern an overall
pattern in the colors. In the bluest filters the solar metallicity
model is much redder at maximum light, until we reach the F435W
-F475W color, where the solar metallicity model suddenly becomes bluer
than the lower metallicity model, redward there is no strong trend in
the colors. Prior to maximum light, there is really no strong trend in
the colors at any wavelength. The largest discrepancy is in fact evident from
Figure~\ref{fig:metal_comp} where the solar metallicity model is
clearly much brighter in the $U-$band. In fact, the solar metallicity
model is nearly 0.5~mag brighter than the low metallicity model in
$U$, even though both models are blue in $U-B$ and the solar
metallicity model is only about 0.1~mag bluer in $U-B$ than the low
metallicity model. Thus, one really needs spectra in the restframe
$U-$band to try to get a handle on progenitor metallicity. 

Overall, the results make sense, in the bluest colors, particularly at
maximum light, the solar metallicity model is much redder than the
lower metallicity model due to higher opacities in the UV, but clear
diagnostics are not so evident.
Our results
are in good agreement with those of Hoeflich  and collaborators
\citep{hwt98,hnuw00}.
\citet{mazz11fe14} found evidence for a progenitor metallicity of
$Z_\odot/2$, which is not inconsistent with our results.

\begin{figure}
\includegraphics[scale=0.4]{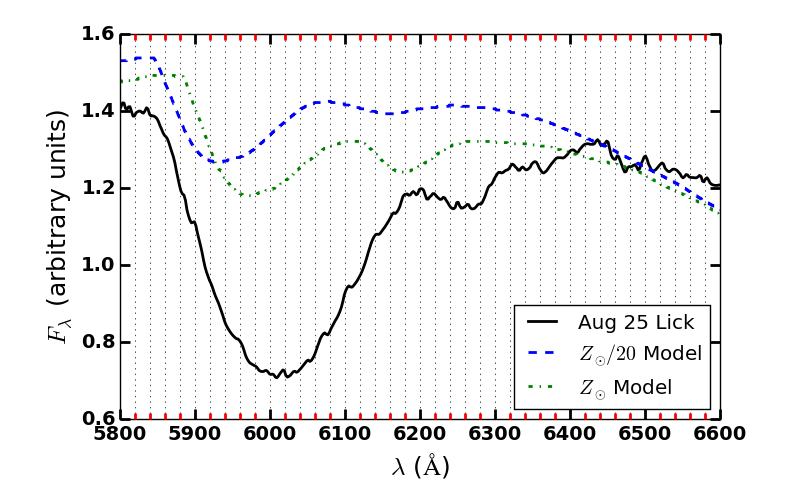}
\caption{The synthetic spectra of the $Z = Z_\odot$ (dashed line) and
  the $Z = Z_\odot/20$ (dot-dashed line) progenitor model compared to
  the observation (solid line) on
  Aug 25. While both models show a line due to C~II $\lambda6580$, the
  solar metallicity model line is about the right shape and strength as
  the observed feature. However the velocity minima are at
  15,000~\kmps in the observations, 18,000 \kmps in the solar
  metallicity model and 19,000 \kmps in the model with $Z = Z_\odot/20$.}
\label{fig:metal_comp_carbon}
\end{figure}

Figure~\ref{fig:metal_comp_carbon} shows that the C~II $\lambda 6580$ appears
with roughly the strength and shape as does the observed feature
in the solar metallicity model at on Aug 25, while not at 
the correct Doppler shift.
The velocities of the feature are
$(15,000,18,000,19,000)$~\kmps for the observations, the solar
metallicity progenitor, and the low metallicity progenitor,
respectively. We have confirmed the identification of the line as due
to C~II in the models by rerunning the spectra with the C~II line
opacity set to zero. This gives
credence to claims of C~II in other SNe~Ia \citep[see][and references
therein]{rollin_Cii11,jerod_cii11,thomas06d07,folatelli12,parrent_screed15}.

Weaker C~II and higher Si velocities can be understood in terms of
the variation in the progenitor evolution. At the end of the 
central He-burning and  therefore, in mixtures which are helium  poor,
$^{12}C(\alpha, \gamma) ^{16}O $ results in  
lower C/O ratios in the convective core. With lower metallicity, the
size of the convective core
increases due to  lower opacities. Shell burning will produce material
with $C/O \approx 1$.
As a result of less carbon, the explosion energy for our progenitors
decreases with $Z$. 
With higher explosion energy, we expand faster and for a given DDT,
the very outer layers  
are burned at higher density reducing the amount of material in the
unburned layers and resulting 
in a higher expansion velocities \citep{hwt98,hnuw00,hoeflich10}.

\subsection{Enhanced Nickel  Mixing?}
\label{subsec:nimix}

\begin{figure}
\includegraphics[scale=0.4]{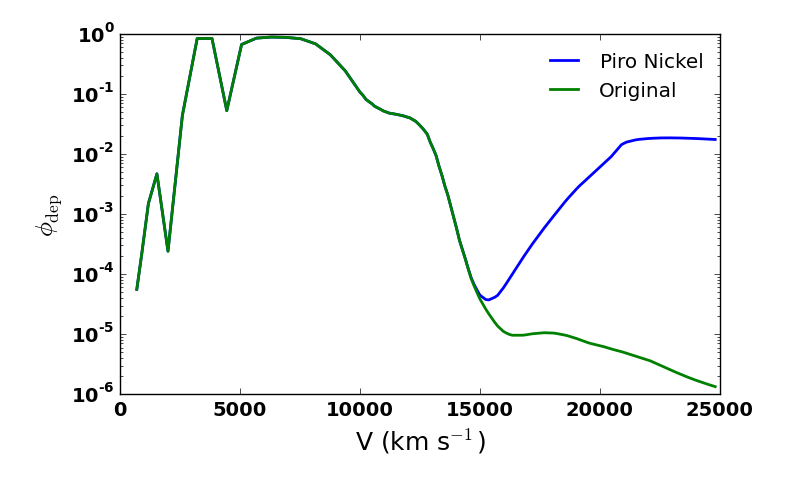}
\caption{The deposition function at day 02, using the model nickel
  distribution and the one obtained by using the prescription of
  \citet{piro11fe12}.} 
\label{fig:piro_gamdep_func}
\end{figure}

We examine a suggestion of \citet{piro11fe12} who by studying
the early rise of the light curve concluded that \nni was required in
the outer $0.1 < M < 1 \times 10^{-3}$ of the supernova. He further
concluded that this requirement places constraints on the
explosion model, somewhat favoring models where the detonation begins
in the outer parts of the star.  Piro's \citeyear{piro11fe12} study is
based on the diffusion time scales for the rise time. 
In fact, geometrical dilution will be responsible for the receding of the
photosphere and adiabatic expansion of the corresponding layers must
be taken into account. Figure~\ref{fig:piro_gamdep_func}
shows the $\gamma$-ray deposition function \citep[the $\gamma$-ray
deposition function is the ratio of the rate of absorption of
$\gamma$-rays $\kappa J$ to the instantaneous decay rate of \nni, see,
for example,][]{sw84,jeff98} for our standard case,
compared to that obtained using the prescription of
\citet{piro11fe12}. 
We chose to make the mass
fraction of radioactive nickel constant at $X_{^{56}Ni} = 0.03$ above a
velocity of 15,500 km/s, which corresponds to a mass of 0.1~\msol
masses from the outer boundary. This is about in the middle of the
parameter 
choices outlined in \citet{piro11fe12}.
In our model the bottom layer where the \nni is
added occurs at a velocity of $v \sim 15,500$~\kmps. Thus, the
photosphere in our model is significantly  above this modified heating
region. Nevertheless, we find negligible changes due to the added
heating. The temperature of our models is a little higher in the outer
parts, but not enough higher in the low density environment to
significantly alter the observed spectra. 
Figure~\ref{fig:piro_gamdep_func} explains this result: the
$\gamma$-ray deposition  function $\phi_\mathrm{dep}$ is increased by
the extra nickel only up to the level of $\sim 1 \times 10^{-3}$, but
only at the highest velocities and not near the photosphere. The
increase in deposition is not
enough to dramatically change the outer energy deposition rate at Day
02.  

\subsection{Comparison to other work}
Light curves and spectra of SN~2011fe at these epochs have
  been studied by \citet{roepke11fe12}, \citet{foley12c}, \citet{FK13},
  \citet{dessart11fe14}, 
  \citet{mazz11fe14}, and \citet{Graham11fe_11by15}. \citet{roepke11fe12} compared the fidelity of a
  delayed detonation Chandrasekhar mass model and violent merger model
  to the observations of the Nearby Supernova Factory and found that
  both models had their shortcomings, but found no strong preference
  for either model. \citet{mazz11fe14} performed abundance tomography
  on a W7 model and a delayed detonation model and found that the
  delayed detonation model was somewhat preferable, in addition to
  finding a primordial metallicity that was about half
  solar. \citet{dessart11fe14} studied PDD models in general, and
  found a preference for a PDD model to match the B-band light curve
  of SNe 2011fe and the spectral evolution. However, their PDD models
  have much weaker shells then the model we considered (compare their
  Figure 10 to Figure~\ref{fig:rho_orig_rho_new}). They also
  find very washed out features at early epochs (see their Figure 16).
  \cite{FK13} used empirical comparisons with the models of
  \citet{lentzmet00} to find a low metallicity for SN~2011fe of about
  $Z_\odot/4$ and a ratio between the metalicity of SN~2011by and
  SN~2011fe of about 30. \citet{Graham11fe_11by15} found the that UV
  flux in the assumed supersolar SN~2011by is significantly lower than
  that in SN~2011fe, in rough agreement with the W7-based models of
  \citet{lentzmet00}, but in contradistinction to the DD models used
  in this work.
Thus, overall our results are generally compatible with those of
previous work.

\section{Conclusions}

We have compared \emph{HST} and ground based spectra of SN~2011fe up
to maximum light to detailed NLTE
synthetic spectra of a Branch-normal hydrodynamic model.
Progenitor models with metallicities of  of $Z_\odot$ and $Z_\odot/20$ 
qualitatively show overall reasonable results after day 2  
based on optical spectra. 

We explored the possibility that the high velocity feature reported by
\citet{jerod11fe12}  
may be produced by a shell as a result of a ``low amplitude pulsation''.
Though a shell may produce  a high velocity feature, it also would produce
high fluxes in the UV and U wavelengths inconsistent with the
observations as well as the very high velocity photosphere at early
times, which leads to washed out features \citep[\S~\ref{sec:results} and][]{dessart11fe14}.
Branch normal models show
features at the corresponding wavelengths produced by iron-group elements
\citep{gerardy03du04}.

 Nearly all the features in the observed spectra are seen in the
 synthetic spectra in approximately the
correct range of the velocity, that is, the elemental
stratification in the model also appears in the observations.
 However, as discussed in \S\ref{sec:res:bench}, the photosphere in
 the model is
 formed at too high  
velocities at early times corresponding to the outermost 0.05 to 0.1 \msol.
 In principle, this can be corrected by a smaller kinetic
energy of the explosion, e.g., by higher central densities and, thus, higher
binding energy, or high C/O ratios in the progenitor.  Alternatively,
the photosphere of SN~2011fe may recede faster due to lower excitation
of ions
(from gamma-rays or the radiation field) 
and/or
variations in the density structure. In SN~2011fe, the second
explanation is favored because, by maximum light, the photospheric velocities
of the model and SN~2011fe are in close agreement (see Table~\ref{tab:vphot}).

There are a multiple ways that the adopted model could be adjusted
to reproduce the observed discrepancies,
including variations in the chemical structure,
rotation of the progenitor, or pulsations prior and during the
explosion (although this would have to done in accord with constraints
from observed spectra).  
Future studies will examine these  effects in detail.

\section*{Acknowledgments}
We thank Aaron Dotter for help
in constructing the synthetic photometry with a wide choice of filters.
We thank the anonymous referee for improving the presentation of this work.
The work has been supported in part by
support for programs
HST-GO-12298.05-A, and HST-GO-12948.04-A   was provided by NASA through a grant from the
Space Telescope Science Institute, which is operated by the
Association of Universities for Research in Astronomy, Incorporated,
under NASA contract NAS5-26555.
This work was also supported in part by the NSF, 
AST-0709181, AST-0707704, 
AST-0708855, AST-0708873.
This
research was also supported, in part, by the NSF grant AST-0703902 to
PAH.  The work of EB was also supported in part by SFB 676, GRK 1354
from the DFG.
ID has been supported in part by the Spanish Ministry
of Science and Innovation project AYA2008-04211-C02-02 (ID).
This research used resources of the
National Energy Research Scientific Computing Center (NERSC), which is
supported by the Office of Science of the U.S.  Department of Energy
under Contract No.  DE-AC02-05CH11231; and the H\"ochstleistungs
Rechenzentrum Nord (HLRN).  We thank both these institutions for a
generous allocation of computer time.

\bibliography{apj-jour,mystrings,refs,baron,sn1bc,sn1a,sn87a,snii,stars,rte,cosmology,gals,agn,atomdata,crossrefs}

\begin{thebibliography}{61}
\expandafter\ifx\csname natexlab\endcsname\relax\def\natexlab#1{#1}\fi

\bibitem[{{Argast} {et~al}\mbox{.}(2001){Argast}, {Samland}, {Gerhard}, \&
  {Thielemann}}]{argast_rev01}
{Argast} D., {Samland} M., {Gerhard} O.~E., {Thielemann} F.-K., 2001,
  Astrophysics and Space Science Supplement, 277, 193

\bibitem[{Baron {et~al}\mbox{.}(2006)Baron, Bongard, Branch, \&
  Hauschildt}]{bbbh06}
Baron E., Bongard S., Branch D., Hauschildt P., 2006, ApJ, 645, 480

\bibitem[{Baron, Branch \& Hauschildt(2007)Baron, Branch, \&
  Hauschildt}]{bbh07}
Baron E., Branch D., Hauschildt P.~H., 2007, ApJ, 662, 1148

\bibitem[{Baron \& Hauschildt(1998)}]{bhpar298}
Baron E., Hauschildt P.~H., 1998, ApJ, 495, 370

\bibitem[{Baron {et~al}\mbox{.}(1996)Baron, Hauschildt, Nugent, \&
  Branch}]{snefe296}
Baron E., Hauschildt P.~H., Nugent P., Branch D., 1996, MNRAS, 283, 297

\bibitem[{{Baron} {et~al}\mbox{.}(2012){Baron}, {Hoeflich}, {Krisciunas},
  {Dominguez}, {Khokhlov}, {Phillips}, {Suntzeff}, \& {Wang}}]{b01ay}
{Baron} E., {Hoeflich} P., {Krisciunas} K., {Dominguez} I., {Khokhlov} A.~M.,
  {Phillips} M.~M., {Suntzeff} N., {Wang} L., 2012, ApJ, 753, 105

\bibitem[{{Bloom} {et~al}\mbox{.}(2012){Bloom} {et~al.}}]{bloom11fe12}
{Bloom} J.~S., {et~al.}, 2012, \apjl, 744, L17

\bibitem[{Bongard {et~al}\mbox{.}(2008)Bongard, Baron, Smadja, Branch, \&
  Hauschildt}]{bongard08}
Bongard S., Baron E., Smadja G., Branch D., Hauschildt P., 2008, ApJ, 687, 456

\bibitem[{{Branch} {et~al}\mbox{.}(2005){Branch}, {Baron}, {Hall}, {Melakayil},
  \& {Parrent}}]{branchcomp105}
{Branch} D., {Baron} E., {Hall} N., {Melakayil} M., {Parrent} J., 2005, PASP,
  117, 545

\bibitem[{{Branch}, {Dang} \& {Baron}(2009){Branch}, {Dang}, \&
  {Baron}}]{branchcomp509}
{Branch} D., {Dang} L.~C., {Baron} E., 2009, PASP, 121, 238

\bibitem[{{Branch} {et~al}\mbox{.}(2006){Branch}, {Dang}, {Hall}, {Ketchum},
  {Melakayil}, {Parrent}, {Troxel}, {Casebeer}, {Jeffery}, \&
  {Baron}}]{branchcomp206}
{Branch} D. {et~al.}, 2006, PASP, 118, 560

\bibitem[{{Branch} {et~al}\mbox{.}(2008){Branch}, Jeffery, Parrent, Baron,
  Troxel, Stanishev, Keithly, Harrison, \& Bruner}]{branch_post08}
---, 2008, PASP, 120, 135

\bibitem[{{Branch} {et~al}\mbox{.}(2007){Branch} {et~al.}}]{branch_pre07}
{Branch} D., {et~al.}, 2007, PASP, 119, 709

\bibitem[{{Brown} {et~al}\mbox{.}(2012){Brown}, {Dawson}, {de Pasquale},
  {Gronwall}, {Holland}, {Immler}, {Kuin}, {Mazzali}, {Milne}, {Oates}, \&
  {Siegel}}]{brown11fe12}
{Brown} P.~J. {et~al.}, 2012, \apj, 753, 22

\bibitem[{{Chomiuk} {et~al}\mbox{.}(2012){Chomiuk} {et~al.}}]{chomiuk11fe12}
{Chomiuk} L., {et~al.}, 2012, \apj, 750, 164

\bibitem[{{Dessart} {et~al}\mbox{.}(2014){Dessart}, {Blondin}, {Hillier}, \&
  {Khokhlov}}]{dessart11fe14}
{Dessart} L., {Blondin} S., {Hillier} D.~J., {Khokhlov} A., 2014, \mnras, 441,
  532

\bibitem[{Doull \& Baron(2011)}]{doull11}
Doull B., Baron E., 2011, PASP, 123, 765

\bibitem[{{Folatelli} {et~al}\mbox{.}(2012){Folatelli} {et~al.}}]{folatelli12}
{Folatelli} G., {et~al.}, 2012, \apj, 745, 74

\bibitem[{{Foley}(2012)}]{foley12c}
{Foley} R.~J., 2012, \apj, 748, 127

\bibitem[{{Foley} \& {Kirshner}(2013)}]{FK13}
{Foley} R.~J., {Kirshner} R.~P., 2013, \apjl, 769, L1

\bibitem[{{Foley} {et~al}\mbox{.}(2012{\natexlab{a}}){Foley}
  {et~al.}}]{foley12a}
{Foley} R.~J., {et~al.}, 2012{\natexlab{a}}, \aj, 143, 113

\bibitem[{{Foley} {et~al}\mbox{.}(2012{\natexlab{b}}){Foley}
  {et~al.}}]{foley12b}
---, 2012{\natexlab{b}}, \apj, 752, 101

\bibitem[{Gerardy {et~al}\mbox{.}(2003)Gerardy {et~al.}}]{gerardy03du04}
Gerardy C., {et~al.}, 2003, ApJ, 607, 391

\bibitem[{{Graham} {et~al}\mbox{.}(2015){Graham}, {Foley}, {Zheng}, {Kelly},
  {Shivvers}, {Silverman}, {Filippenko}, {Clubb}, \&
  {Ganeshalingam}}]{Graham11fe_11by15}
{Graham} M.~L. {et~al.}, 2015, \mnras, 446, 2073

\bibitem[{Hauschildt \& Baron(1999)}]{hbjcam99}
Hauschildt P.~H., Baron E., 1999, J. Comp. Applied Math., 109, 41

\bibitem[{Hauschildt, Baron \& Allard(1997)Hauschildt, Baron, \&
  Allard}]{hbapara97}
Hauschildt P.~H., Baron E., Allard F., 1997, ApJ, 483, 390

\bibitem[{Hauschildt {et~al}\mbox{.}(1996)Hauschildt, Baron, Starrfield, \&
  Allard}]{phhnovfe296}
Hauschildt P.~H., Baron E., Starrfield S., Allard F., 1996, ApJ, 462, 386

\bibitem[{Hauschildt {et~al}\mbox{.}(1997)Hauschildt, Schwarz, Baron,
  Starrfield, Shore, \& Allard}]{phhnovetal97}
Hauschildt P.~H., Schwarz G., Baron E., Starrfield S., Shore S., Allard F.,
  1997, ApJ, 490, 803

\bibitem[{{Hoeflich}(2002)}]{h02}
{Hoeflich} P., 2002, New Astronomy Review, 46, 475

\bibitem[{{Hoeflich}(2006)}]{hoef_review06}
---, 2006, Nuclear Physics A, 777, 579

\bibitem[{Hoeflich {et~al}\mbox{.}(2002)Hoeflich, Gerardy, Fesen, \&
  Sakai}]{HGFS99by02}
Hoeflich P., Gerardy C., Fesen R., Sakai S., 2002, ApJ, 568, 791

\bibitem[{Hoeflich, Khokhlov \& Wheeler(1995)Hoeflich, Khokhlov, \&
  Wheeler}]{hofdd+mol95}
Hoeflich P., Khokhlov A., Wheeler J.~C., 1995, ApJ, 444, 831

\bibitem[{{Hoeflich} {et~al}\mbox{.}(2010){Hoeflich}, {Krisciunas}, {Khokhlov},
  {Baron}, {Folatelli}, {Hamuy}, {Phillips}, {Suntzeff}, \&
  {Wang}}]{hoeflich10}
{Hoeflich} P. {et~al.}, 2010, ApJ, 710, 444

\bibitem[{{Hoeflich}, {M{\"u}ller} \& {Khokhlov}(1993){Hoeflich}, {M{\"u}ller},
  \& {Khokhlov}}]{hmk93b}
{Hoeflich} P., {M{\"u}ller} E., {Khokhlov} A., 1993, A\&A, 268, 570

\bibitem[{{Hoeflich} {et~al}\mbox{.}(2000){Hoeflich}, {Nomoto}, {Umeda}, \&
  {Wheeler}}]{hnuw00}
{Hoeflich} P., {Nomoto} K., {Umeda} H., {Wheeler} J.~C., 2000, \apj, 528, 590

\bibitem[{Hoeflich, Wheeler \& Thielemann(1998)Hoeflich, Wheeler, \&
  Thielemann}]{hwt98}
Hoeflich P., Wheeler J.~C., Thielemann F.-K., 1998, ApJ, 495, 617

\bibitem[{{Horesh} {et~al}\mbox{.}(2012){Horesh} {et~al.}}]{horesh11fe12}
{Horesh} A., {et~al.}, 2012, \apj, 746, 21

\bibitem[{Hsiao {et~al}\mbox{.}(2013)Hsiao {et~al.}}]{hsiao11fe13}
Hsiao E., {et~al.}, 2013, \apj, 766, 72

\bibitem[{{Isern} {et~al}\mbox{.}(2013){Isern} {et~al.}}]{isern11fe13a}
{Isern} J., {et~al.}, 2013, \aap, 552, A97

\bibitem[{{Jeffery}(1998)}]{jeff98}
{Jeffery} D.~J., 1998, ArXiv Astrophysics e-prints, astro-ph/9811356

\bibitem[{Lentz {et~al}\mbox{.}(2000)Lentz, Baron, Branch, Hauschildt, \&
  Nugent}]{lentzmet00}
Lentz E., Baron E., Branch D., Hauschildt P.~H., Nugent P., 2000, ApJ, 530, 966

\bibitem[{{Li} {et~al}\mbox{.}(2011){Li} {et~al.}}]{weidong_last}
{Li} W., {et~al.}, 2011, \nat, 480, 348

\bibitem[{{Liu} {et~al}\mbox{.}(2012){Liu}, {Di Stefano}, {Wang}, \&
  {Moe}}]{liu11fe12}
{Liu} J., {Di Stefano} R., {Wang} T., {Moe} M., 2012, \apj, 749, 141

\bibitem[{{Matheson} {et~al}\mbox{.}(2012){Matheson} {et~al.}}]{matheson11fe12}
{Matheson} T., {et~al.}, 2012, \apj, 754, 19

\bibitem[{Mazzali {et~al}\mbox{.}(2014)Mazzali {et~al.}}]{mazz11fe14}
Mazzali P.~A., {et~al.}, 2014, MNRAS, 439, 1959

\bibitem[{{Munari} {et~al}\mbox{.}(2013){Munari}, {Henden}, {Belligoli},
  {Castellani}, {Cherini}, {Righetti}, \& {Vagnozzi}}]{munari11fe13}
{Munari} U., {Henden} A., {Belligoli} R., {Castellani} F., {Cherini} G.,
  {Righetti} G.~L., {Vagnozzi} A., 2013, \na, 20, 30

\bibitem[{{Nugent} {et~al}\mbox{.}(2011){Nugent} {et~al.}}]{nug_11fe_11}
{Nugent} P.~E., {et~al.}, 2011, Nature, 480, 344

\bibitem[{{Ohlmann} {et~al}\mbox{.}(2014){Ohlmann}, {Kromer}, {Fink}, {Pakmor},
  {Seitenzahl}, {Sim}, \& {R{\"o}pke}}]{ohlmann14}
{Ohlmann} S.~T., {Kromer} M., {Fink} M., {Pakmor} R., {Seitenzahl} I.~R., {Sim}
  S.~A., {R{\"o}pke} F.~K., 2014, \aap, 572, A57

\bibitem[{{Parrent}(2015)}]{parrent_screed15}
{Parrent} J.~T., 2015, \mnras, {submitted astro-ph/1412.7163}

\bibitem[{{Parrent} {et~al}\mbox{.}(2011){Parrent}, {Thomas}, {Fesen},
  {Marion}, {Challis}, {Garnavich}, {Milisavljevic}, {Vink{\`o}}, \&
  {Wheeler}}]{jerod_cii11}
{Parrent} J.~T. {et~al.}, 2011, \apj, 732, 30

\bibitem[{{Parrent} {et~al}\mbox{.}(2012){Parrent} {et~al.}}]{jerod11fe12}
{Parrent} J.~T., {et~al.}, 2012, \apjl, 752, L26

\bibitem[{{Pereira} {et~al}\mbox{.}(2013){Pereira} {et~al.}}]{factory11fe13}
{Pereira} R., {et~al.}, 2013, \aap, 554, A27

\bibitem[{{Piro}(2012)}]{piro11fe12}
{Piro} A.~L., 2012, \apj, 759, 83

\bibitem[{Quimby {et~al}\mbox{.}(2006)Quimby, Hoeflich, Kannappan, Rykoff,
  Rujopakarn, Akerlof, Gerardy, \& Wheeler}]{quimby05cg06}
Quimby R., Hoeflich P., Kannappan S., Rykoff E., Rujopakarn W., Akerlof C.,
  Gerardy C., Wheeler J.~C., 2006, ApJ, 636, 400

\bibitem[{{Richmond} \& {Smith}(2012)}]{RS11fe12}
{Richmond} M.~W., {Smith} H.~A., 2012, Journal of the American Association of
  Variable Star Observers (JAAVSO), 40, 872

\bibitem[{{R{\"o}pke} {et~al}\mbox{.}(2012){R{\"o}pke} {et~al.}}]{roepke11fe12}
{R{\"o}pke} F.~K., {et~al.}, 2012, \apjl, 750, L19

\bibitem[{Sandage {et~al}\mbox{.}(1996)Sandage, Saha, Tammann, Labhardt,
  Panagia, \& Macchetto}]{sandage90N96}
Sandage A., Saha A., Tammann G., Labhardt L., Panagia N., Macchetto F.~D.,
  1996, ApJ, 460, L15

\bibitem[{Sutherland \& Wheeler(1984)}]{sw84}
Sutherland P., Wheeler J.~C., 1984, ApJ, 280, 282

\bibitem[{{Thomas} {et~al}\mbox{.}(2007){Thomas} {et~al.}}]{thomas06d07}
{Thomas} R.~C., {et~al.}, 2007, ApJ, 654, L53

\bibitem[{{Thomas} {et~al}\mbox{.}(2011){Thomas} {et~al.}}]{rollin_Cii11}
---, 2011, \apj, 743, 27

\bibitem[{{Vink{\'o}} {et~al}\mbox{.}(2012){Vink{\'o}}, {S{\'a}rneczky},
  {Tak{\'a}ts}, {Marion}, {Heged{\"u}s}, {B{\'{\i}}r{\'o}}, {Borkovits},
  {Szegedi-Elek}, {Farkas}, {Klagyivik}, {Kiss}, {Kov{\'a}cs}, {P{\'a}l},
  {Szak{\'a}ts}, {Szalai}, {Szalai}, {Szatm{\'a}ry}, {Szing}, {Vida}, \&
  {Wheeler}}]{vinko11fe12}
{Vink{\'o}} J. {et~al.}, 2012, \aap, 546, A12

\end{thebibliography}

\bsp

\label{lastpage}

\end{document}

p